# Effective climate policies for major emission reductions of ozone precursors: Global evidence from two decades

Ningning Yao[1,2,8], Huan Xi[1,2,8], Lang Chen[1,2,8], Zhe Song[3,8], Jian Li[2], Yulei Chen[2], Baocai Guo[1], Yuanhang Zhang[4], Tong Zhu[4], Pengfei Li[5*], Daniel Rosenfeld[6], and John H. Seinfeld[7], and Shaocai Yu[2*]

[1]Collaborative Innovation Center for Statistical Data Engineering, Technology and Application; School of Statistics and Mathematics, Zhejiang Gongshang University, Hangzhou, 310018, China

[2]Zhejiang Province Key Laboratory of Solid Waste Treatment and Recycling; School of Environmental Sciences and Engineering, Zhejiang Gongshang University, Hangzhou, 310018, China

[3]Research Center for Air Pollution and Health, Key Laboratory of Environmental Remediation and Ecological Health, Ministry of Education, College of Environment and Resource Sciences, Zhejiang University, Hangzhou, 310058, China

[4] College of Environmental Sciences and Engineering, Peking University, Beijing 100871, China

[5] State Key Laboratory of Infrared Physics, Shanghai Institute of Technical Physics, Chinese Academy of Sciences, Shanghai, 200031, China

[6] Institute of Earth Sciences, The Hebrew University of Jerusalem, Jerusalem, Israel

[7] Division of Chemistry and Chemical Engineering, California Institute of Technology, Pasadena, CA 91125, USA

[8]These authors contributed equally: Ningning Yao, Huan Xi, Lang Chen, Zhe Song.

[*]To whom correspondence may be addressed. Email: shaocaiyu@zjgsu.edu.cn; pengfeili@mail.sitp.ac.cn

**Abstract:** Despite policymakers deploying various tools to mitigate emissions of $O_3$ precursors, such as nitrogen oxides ($NO_x$), carbon monoxide (CO) and volatile organic compounds (VOCs), the effectiveness of policy combinations remains uncertain. We employ an integrated framework that couples structural break detection with machine learning to pinpoint effective interventions across the building, electricity, industrial and transport sectors, identifying treatment effects as abrupt changes without prior assumptions about policy treatment assignment and timing. Applied to two decades of global $O_3$ precursors emissions data, we detect 78, 77 and 78 structural breaks for $NO_x$, CO and VOCs, corresponding to cumulative emission reductions of 0.96–0.97 Gt, 2.84–2.88 Gt, and 0.47–0.48 Gt, respectively. Sector-level analysis shows electricity-sector structural policies cut $NO_x$ by up to 32.4 %, while in buildings, developed countries combined adoption subsidies with carbon taxes to achieve 42.7 % CO reductions and developing countries used financing plus fuel taxes to secure 52.3 %. VOCs abatement peaked at 38.5 % when fossil-fuel subsidy reforms were paired with financial incentives. Finally, hybrid strategies merging non-price measures (subsidies, bans, mandates) with pricing instruments delivered up to an additional 10 % co-benefit. These findings guide the sequencing and complementarity of context-specific policy portfolios for $O_3$ precursor mitigation.

# Introduction

Tropospheric ozone ($O_3$), as a short-lived pollutant, forms through photochemical reactions involving nitrogen oxides ($NO_x$), carbon monoxide (CO), and volatile organic compounds (VOCs), while elevated regional $O_3$ concentrations can significantly alter the surface radiation balance, with local warming effects up to several hundred times that of carbon dioxide[1-3]. This pollution not only threatens human health and agricultural productivity but also creates a positive feedback loop with climate change. An increase of 1°C in temperature can enhance photochemical reaction rates by 3–5%, while intensified extreme heat events boost VOCs emissions from vegetation, leading to a "high temperature–$O_3$" coupling effect[4,5]. The climatic impacts of $O_3$ precursors are multifaceted: VOCs and $NO_x$ contribute to the formation of secondary organic aerosols, which affect solar radiation absorption and cloud albedo, thereby intensifying radiative forcing. Regional observations indicate that their impacts on solar radiation weakening are 2–3 times greater than those of natural background levels[6,7]. Additionally, increased atmospheric oxidative capacity accelerates the conversions of methane and other greenhouse gases, shortening methane's atmospheric lifetime by 10–15%, but promoting its conversion to carbon dioxide ($CO_2$), thus posing long-term climate risks[8].

International experience demonstrates that coordinated pollution abatement and carbon mitigation can yield synergistic benefits. Market-based climate instruments such as carbon taxes and emissions trading schemes have been shown to directly curb $NO_x$ emissions and, through altered fuel use and technological shifts, to deliver indirect VOCs reductions, provided that their transitional design is tailored to regional emission sources and economic structures[9-13]. Consequently, Policymakers have thus deployed a broad toolkit—from command-and-control standards targeting vehicles, fuels and industrial stacks to economic instruments (carbon or pollution taxes, tradable permits) and technology mandates or subsidies for clean energy and enhanced efficiency[14,15]. Theoretically, many of these measures should confer air quality co-benefits, However, few studies have systematically quantified the interactions among multiple policy levers[16-18], leaving empirical evidence on their combined impacts, particularly concerning key $O_3$ precursors such as $NO_x$, CO, and VOCs—fragmented and insufficient[19]. Moreover, global assessments of $O_3$ precursor emissions are constrained by data scarcity and model limitations. While the Organization for Economic Co-operation and Development (OECD) provides high-quality policy data, coverage gaps persist, hindering the comparisons of cross-national policy effects[20,21]. Climate change further complicates $O_3$ management by intensifying natural precursor emissions, altering atmospheric dynamics, and prolonging pollutant residence times, thereby heightening the uncertainty surrounding traditional abatement measures[22,23]. European studies indicate that warming trends may erode the long‐term gains from emission reduction policies, with projected $O_3$ increases in certain regions offsetting past improvements[24,25]. Crucially, existing policy evaluations typically hinge on researcher-selected intervention scenarios, risking the exclusion of vital policy instruments and obscuring the true scope of synergistic effects under a changing climate[26-28].

Addressing these challenges requires a systematic approach that integrates data consolidation, technological innovation, and policy synergy. The proposed strategies include establishing a collaborative observation system covering multiple sectors to create a dynamic global database of $O_3$ precursor emissions[29,30]. Utilizing machine learning models can help decipher the complex relationships between policy interventions and emission changes[31], facilitating dynamic and accurate policy assessments. Developing standardized assessment frameworks that encompass various policy tools enables uniform standards for evaluating the effectiveness of cross-border policies[32, 33], from qualitative regional descriptions to quantitative global analyses by intergovernmental bodies such as the Intergovernmental Panel on Climate Change (IPCC).

This study aims to conduct a large-scale causal impact assessment based on global data to identify which climate policies implemented by countries around the world over the past 20 years have effectively reduced $O_3$ precursor emissions. By consistently and theoretically classifying various climate policy instruments, the synergistic effects were systematically evaluated among different instruments[34]. However, in global climate governance and policy evaluation, a major empirical challenge is the extensive policy interventions in the real world. Due to statistical degrees of freedom constraints, traditional methods like difference-in-differences (DID) and two-way fixed effects (TWFE) models face limitations when dealing with extensive policy interventions and complex policy mixes. Additionally, existing methods primarily focus on verifying known policy causality, making it difficult to address broader reverse causality issues, especially in situations where intervention timings and mechanisms are unclear[35, 36]. To deal with such problems, the extended two-way fixed effect (ETWFE) model flexibly captures the dynamic impact of heterogeneous processing effects and overlapping processing by introducing the interaction term between processing queue and time[37]. However, ETWFE still depends on predefined treatment variables, making it insufficient to identify unspecified interventions. In response, Pretis and Schwarz proposed a structural break-driven reverse causal inference framework that redefined the problem of reverse causality in the context of variable and model selection, with a focus on detecting and estimating the causal effects of unknown interventions, including their timing and allocation mechanisms[38]. Based on these advances, our study integrates the heterogeneous treatment capabilities of the ETWFE framework with structural break detection techniques to address $O_3$ precursor emissions. By introducing the term of time country interaction, we model treatment effects as structural changes within individual fixed effects. By combining machine learning tools to select high-dimensional virtual variables, we automatically pinpoint key intervention times and targets without prior assumptions about policy treatment assignment and timing. We apply the TWFE-DID models to identify which tools or combinations can most significantly reduce $O_3$ precursor emissions. This offers a flexible and robust solution for causal inference of large-scale policy data, particularly suitable for addressing reverse causality problems related to factors that generally reduce emissions.

## Data and Methods

### Data description

This study utilizes the Climate Action and Policy Measurement Framework (CAPMF) dataset developed by the OECD under the International Programme for Action on Climate (IPAC). The CAPMF dataset is designed to support the global inventory under the United Nations Framework Convention on Climate Change (UNFCCC) through regularly monitoring and evaluating climate mitigation policies[39, 40]. The dataset is publicly accessible through the OECD data browser (https://oe.cd/dx/capmf). The CAPMF provides structured and standardized data on climate policies of OECD and G20 countries from 1998 to 2022, based on specific policy tools and definitions shared by the IPCC and UNFCCC (Note: Data coverage for non-OECD/G20 countries is still being expanded). The CAPMF includes policies with explicit climate mitigation goals or proven mitigation effects. For each policy tool, implementation intensity is quantified through a strict index (0-10 point scale) at the sectoral level[41, 42]. Based on this index, we apply a data-driven approach to identify policy adoption and major tightening events in four key sectors (buildings, electricity, industry, and transport) across 41 sample countries. A summary relevant to the present study is presented here. The policy data are the same as those described in Stechemesser et al[34] and include additional policy events from the International Energy Agency (IEA) Policy[43] and Measures Database and the New Climate Institute Climate Policy Database[44]. This includes 113 policies from the transport sector and 39 policies from the buildings sector.

The sectoral emissions data for $O_3$ precursors ($NO_X$, CO, and VOCs) are sourced from the "Global Atmosphere Watch Emissions Database" (EDGAR) v8.1 of the Joint Research Centre of the European Commission[45]. In this study, volatile organic compounds refer to nonmethane volatile organic compounds (NMVOCs), abbreviated as VOCs, but excluding methane. According to the IPCC methodology, we aggregate emissions data based on IPCC sub-sector codes to construct emission series for the buildings, electricity, industry, and transport sectors[46,47] (Supplementary Table 6). This method ensures that only emission sources directly related to policy data are included. The emission sources corresponding to the original EDGAR classification can be considered negligible for buildings and the electricity sectors. For the transport sector, we exclude emissions from sectors with no corresponding policies, such as maritime transportation and civil aviation. For the industry sector, we focus on emissions from processes and combustion in the chemical, non-metallic mineral, and metal industries[48,49]. For sample selection, countries with relatively small total emissions from these four sectors are excluded, including Israel, due to abnormal emission data from 2013-2014. The final samples include 41 countries from six continents (27 developed countries and 14 developing countries). Socio-economic data, such as population and GDP, are sourced from the World Bank's climate database[50, 51]. Degree-days for heating (HDD, base temperature of 16 °C) and cooling (CDD, base temperature of 18 °C) are provided by the IEA[52]. Our analysis focuses explicitly on policies at the national level. However, out of the 41 countries in our sample, 20 are EU member states that have not only made national-level policy efforts but also participate in coordinated EU-wide climate policies. To consider the impact of EU policies in our model, we incorporate multi-country policies observed at the EU level (from our CAPMF database) as known policies (not overfitted) embedded into the initial model of the developed country group. We include symbols of EU policies in Figs. 2 and 3 for readers to consider these when explaining (Supplementary Table 3).

## Methods

Our objective is to identify, by means of an empirical framework, those policies—among approximately 1,500 candidate interventions in our database—that have a demonstrable impact on reducing $O_3$ precursor emissions. The principal empirical challenge lies in the sheer volume of policies: conventional evaluation tools, which typically focus on a single, pre-selected intervention, are ill-suited to this context. The existing studies[35, 36] commonly choose a policy of interest a priori and, after controlling for other socioeconomic determinants, attempt to isolate its effect. This approach can overlook promising but unexamined measures and disregards the possibility that combinations of policies may produce synergistic benefits. To avoid subjective selection of policies that appear promising for analysis, the classic TWFE model was extended on the basis of recent research progress[34, 38, 53]. A machine learning-driven approach was adopted to identify structural breaks in carbon emissions and attribute them to known or unknown policy interventions and standard effect size estimation methods were combined in causal inference to quantify policy impacts. A detailed discussion of the modelling approach was provided in the recent studies[34, 38, 53], and a summary relevant to the present study is presented here. Starting from a standard TWFE-DID model, we construct a country-year panel dataset in which the dependent variables are the logarithms of $NO_x$, CO and VOCs emission levels. We then integrate break-detection techniques into the TWFE framework: abrupt, step-wise shifts in country fixed effects ("breaks") are identified and, in post-hoc analysis, associated with policy interventions. This hybrid approach enables unbiased discovery of both individual and combined policy effects without relying on prior assumptions regarding policy timing or type.

To achieve this goal, we construct a saturated TWFE model for emission break detection, incorporating dummy variables for all policy interventions across all countries (N=41) and potential policy implementation times (T=22). This approach allows for the identification of potential emission policy effects at any country-time combination below:

$$\ln(O_3\ precursor)_{i,t} = \mu_i + \eta_t + \sum_{j=1}^{N}\sum_{s=2}^{T} \tau_{j,s} 1_{i=j,t\geq s} + \beta_1 \ln(gdp_{i,t}) + \beta_2 \ln(gdp_{i,t})^2 + \beta_3 \ln(pop_{i,t}) + \beta_4 \ln(hdd_{i,t}) + \beta_5 \ln(cdd_{i,t}) + \beta_6 eu_{i,t} + \beta_{7,i} t + \varepsilon_{i,t} \quad (1)$$

Where $\ln(O_3\ precursor)$ is the dependent variable representing the natural logarithm of O3 precursor emissions ($NO_x$, CO, VOCs); $u_i$ denotes the fixed effects of a country, capturing time-invariant country characteristics; $\eta_t$ represents group-year fixed effects (grouped by developed and developing countries), controlling for common time shocks; $1_{i=j,t\geq s}$ is a step indicator variable that equals 1 if country $j$ is intervened at time $s$ or later; otherwise, it is 0. The model includes $N\times(T-1)$ potential break parameters $\tau_{j,s}$, covering all country-year combinations. The control variables are specified as follows: economic factors: logarithm of GDP and its squared term (to capture the nonlinear relationship between emissions and economic development); Population factors: logarithm of population size; Climate factors: heating degree days (HDD) and cooling degree days (CDD); EU policy control variables: dummy variables for cross-national policies such as carbon labeling and emissions trading systems; Country time trend: country-specific time trend terms. $\beta_{7,i} t$ to control for long-term linear changes in emissions across countries.

This study employs a high-dimensional variable selection framework to identify significant emission reduction interventions from sparse step break indicators $\tau_{j,s}$ While rigorously controlling false-positive discovery risks[38]. By using a General-to-Specific (Gets) approach implemented via the R packages 'gets' and 'getspanel', our three-stage algorithm operates under the null hypothesis of no treatment effect to probabilistically control spurious break detection[54]. Firstly, step intervention indicators (SIS) are randomly divided into blocks (size=20) for preliminary screening, and then p-value thresholds calibrated with gamma ($\gamma$) are used for combined path-search evaluation(e.g., $\gamma$ =0.01 corresponds to p=1%). Iterative optimization loops eliminate non-significant variables until all retained candidates meet pre-specified significance criteria. To enhance causal identification, the framework integrates a DID design with panel data from untreated control groups, systematically reducing the risk of Type I error while utilizing cross-sectional counterfactual information. The robustness of the method is verified through sensitivity analyses of γ values (0.001), assessments of impulse intervention indicator (IIS), and parameter configuration tests, confirming its stability in balancing model simplicity and statistical rigor under high-dimensional settings.

Based on the data-driven selection of significant intervention measures (determined by their emission reduction effects relative to the control group and untreated cohort), we estimate effect magnitudes for retained step intervention indicators (SIS) by using a sparse TWFE model. The specification is formalized as follows:

$$\ln (\widehat{O_3 \, precursor})_{i,t} = \hat{\mu}_i + \hat{\eta}_t + \sum_{j \in \widehat{\Gamma}_r} \sum_{s \in \widehat{T}_j} \hat{\tau}_{j,s} \mathbf{1}_{(i=j, t \geq s)} + \hat{\beta}_1 \ln(gdp_{i,t}) + \hat{\beta}_2 \ln(gdp_{i,t})^2 + \hat{\beta}_3 \ln(pop_{i,t}) + \hat{\beta}_4 \ln(hdd_{i,t}) + \hat{\beta}_5 \ln(cdd_{i,t}) + \hat{\beta}_6 eu_{i,t} + \hat{\beta}_{7,i} t + \hat{\varepsilon}_{i,t} \quad (2)$$

Where $\Gamma_r$ denotes the set of countries where structural breaks have been detected, $T_j$ represents break time for country $j$, and $\hat{\tau}_{j,s}$ quantifies the magnitude of heterogeneous policy effects of country $j$ at time $s$, capturing asymmetric emission discontinuities. This framework relaxes traditional DID assumptions of synchronous policy adoption by allowing for spatiotemporal heterogeneity in treatment efficacy[35]. To reduce the selection bias of prioritizing large-effect policies during break detection, we enforce a strict significance threshold (γ≤0.01) and apply post-selection bias correction through the Gets algorithm[38]. When TWFE serves as our baseline estimator, robustness is assessed by using generalized synthetic control methods (GSCM).

Given that our sample includes countries at different stages of economic development, we perform subgroup analyses to address potential heterogeneity. According to the United Nations World Economic Situation and Prospects report, we categorize countries into two groups[55]: (1) 27 developed/industrialized economies (e.g., Germany, USA, Japan) are characterized by mature decarbonization systems and technology-driven mitigation pathways, and (2) 14 developing or transition economies (e.g., China, India, Brazil) have delayed policy implementations and high incremental emission potential due to rapid industrialization. This classification helps to gain a deeper understanding of how economic development levels affect the effectiveness of emission reduction policies in different national contexts.

# Results

## 1. Identifying structural breaks and estimating the effects of $O_3$ precursor emission reductions

Analysis of OECD policy data revealed approximately 1,500 instances of policy adoption and tightening over time, with climate policy counts showing a gradual upward trend (Extended Data

Fig. 1a). Policy instruments exhibit significant cross-sectoral and cross-national heterogeneity (Extended Data Fig. 1b), yet their impacts on emissions remain highly uncertain. Leveraging a machine learning-enhanced ETWFE-DID framework, we analyze a global dataset of $O_3$ precursor emissions ($NO_x$, CO, VOCs) across 41 representative countries (including developed and developing economies) and four economic sectors from 2000 to 2021. This identifies 233 structural breaks associated with substantial emission declines (Fig. 1), displaying pronounced sectoral heterogeneity: the transport sector accounted for the most breaks (68), followed by electricity and building sectors (58 each), and the industrial sector (49) (Fig. 1). This aligns with global observations of marked reductions in transport-related $NO_x$, CO, and VOCs emissions since the early 2000s[56, 57].

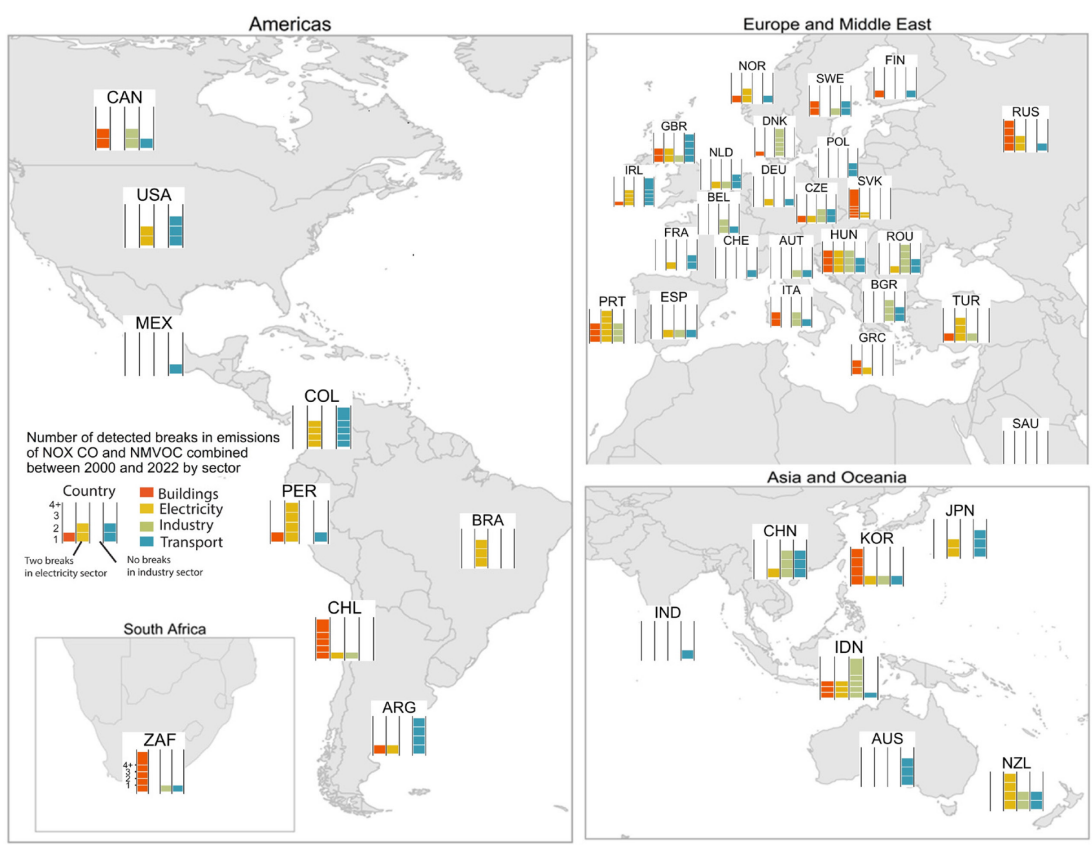

**Fig.1 Break detection analysis of $O_3$ precursor emissions using the TWFE-DID model.** This analysis identifies significant emission reduction events in different countries and economic sectors. A total of 233 significant emission breaks are detected globally (country names are represented by their international ISO codes; **for instance, Bulgaria is denoted as BGR.**), of which 150 breaks are in developed economies and 83 are in developing economies. Developed Economies (27 countries): Australia, Austria, Belgium, Bulgaria, Canada, Czech Republic, Denmark, Finland, France, Germany, Greece, Hungary, Ireland, Italy, Japan, Netherlands, New Zealand, Norway, Poland, Portugal, Romania, Slovakia, Spain, Sweden, Switzerland, United Kingdom, and United States. Developing economies (14 countries): Argentina, Brazil, Chile, China, Colombia, India, Indonesia, Mexico, Peru, Russia, Saudi Arabia, South Africa, South Korea, and Turkey.

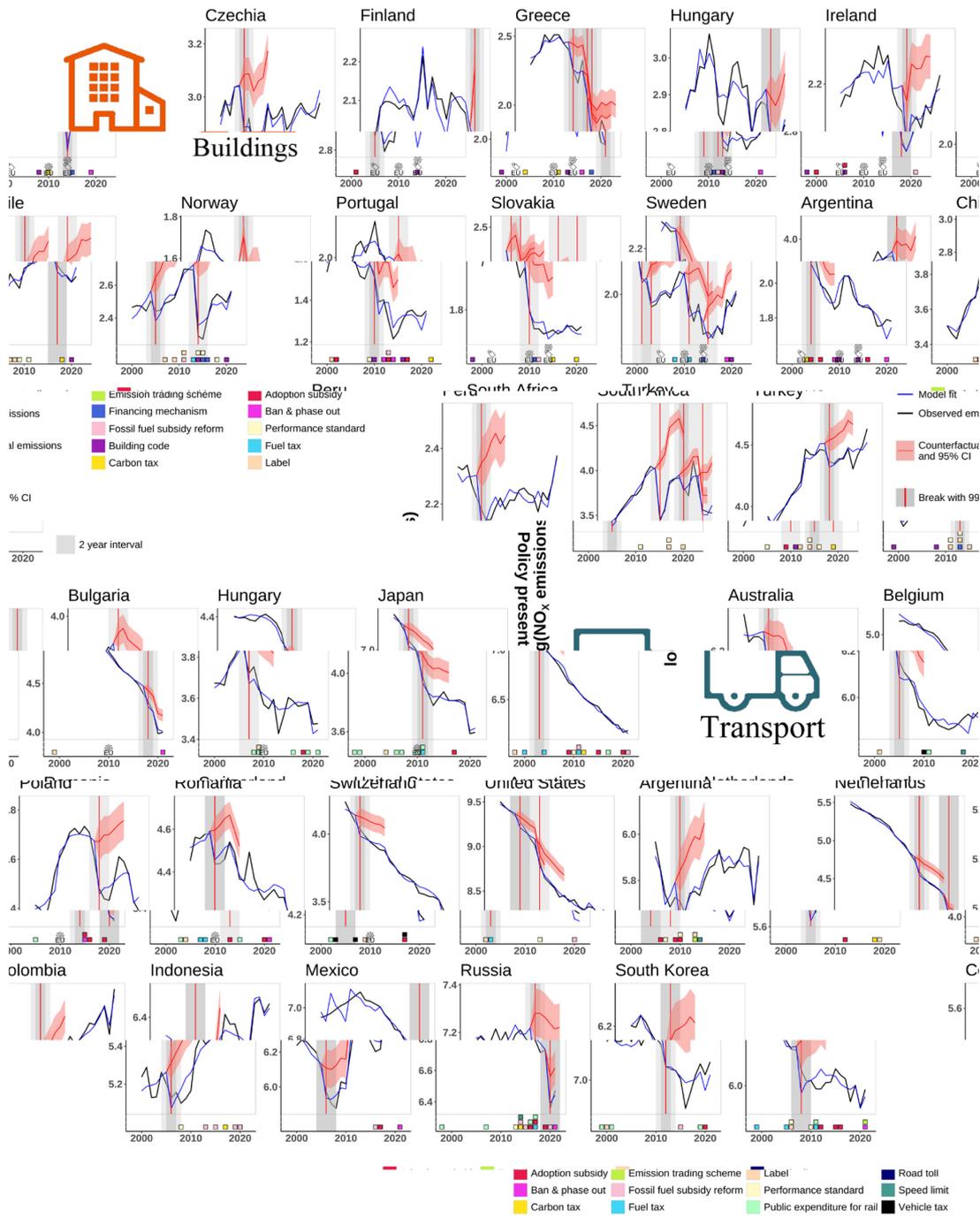

**Fig. 2 Breaks detection and policy matching for NO$_x$ emissions in the buildings and transport sectors.** For each country and sector, the black lines indicate the observed temporal changes in emissions, while the blue lines represent the model's fitting results, which are very consistent with the actual emission data. Detected emission breaks are marked by red vertical lines, while counterfactual emissions are depicted by red lines. Each break is accompanied by a statistical confidence interval (dark gray) and a two-year confidence interval (light gray); These intervals may overlap in some cases. The two-year confidence interval accounts for both statistical uncertainty and potential policy response lags. Policy interventions are denoted by colored squares on the x-axis, representing both newly adopted and tightened policies. The color of each square signifies the type of policy instrument used. If a square falls within the two-year confidence interval of a detected break, the corresponding policy is attributed to that emission break. The adoption and tightening of EU labels, EU performance standards, and the EU emissions trading system are indicated by symbols representing labels, gears, and Euro icons, respectively. In the buildings and transport sectors, 53% and 70% of matched breaks, respectively, are matched with these policy combinations, respectively. Additional analyses for CO and VOC emission breaks and their policy matches are provided in the Extended Data Figs. 2-5.

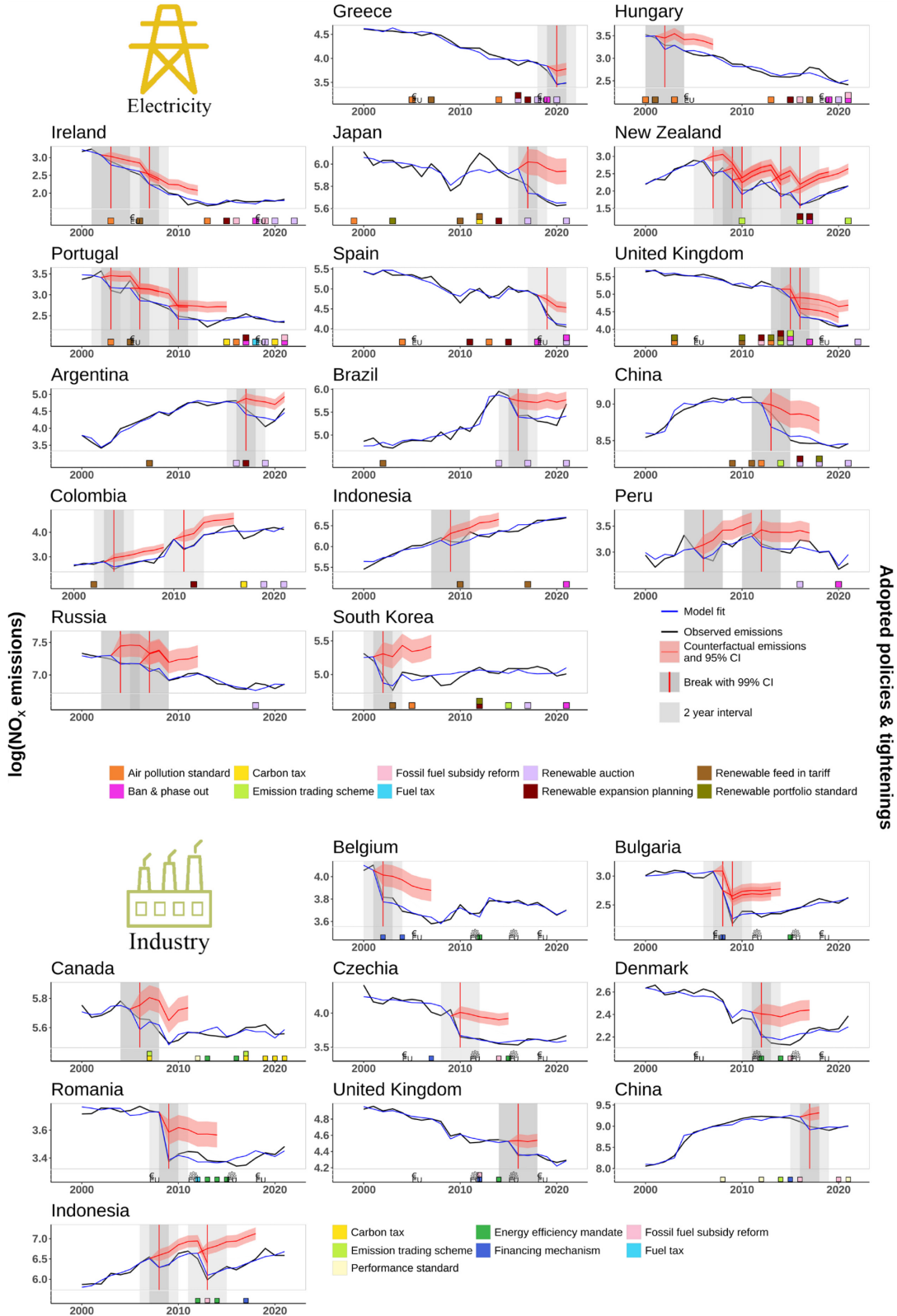

**Fig. 3 The same as Fig. 2 but for the electricity and industrial sectors.** In the electricity and industrial sectors, 53% and 63% of matched breaks, respectively, are matched with these policy combinations.

By using the ETWFE-DID estimator, we quantify emission reductions for each break ($NO_x$ results in Figs. 2–3; CO and VOCs in Extended Data Figs. 2–5). They provide evidence that the detected breaks are credible and align with the timing of the adoption or tightening of meaningful climate policies. Time-series comparisons of observed emissions (black lines in Fig. 2, Fig. 3 and Extended Data Figs. 2–5) and model-predicted emissions (blue lines) demonstrate the model's capacity to capture dynamic relationships between log-transformed precursor emissions and socioeconomic factors (GDP, population), climatic variables (heating/cooling degree days), and country-specific temporal trends. Further comparison with counterfactual emissions (red lines, assuming no policy intervention) confirms significant national-level emission declines at breaks. The summary calculation results show that the number of breaks for $NO_x$, CO, and VOCs is 78, 77, and 78, respectively, and the average emission reduction sizes are 24.2%, 26.9%, and 23.4%, respectively. (Supplementary Tables 1 and 2). We report point estimates and standard errors for country-level emission breaks in Supplementary Tables 7–30. Based on the approximate 95 % confidence intervals of these estimates, we calculate equivalent total emission reductions of $NO_x$ between 0.96 and 0.97 Gt, CO between 2.84 and 2.88 Gt, and VOCs between 0.47 and 0.48 Gt (Supplementary Section 5).

The distributions of emission reductions vary by sector, reflecting differences in policy types and technological characteristics (Supplementary Tables 1 and 2). For $NO_x$ emissions, the electricity sector leads in emission reductions, with 27 breaks (accounting for 50% of sector cases) and the highest average emission reduction effect size (30.1%). In contrast, the average emission reduction effect size of industrial (25.8%) and buildings (22.1%) is lower (Supplementary Tables 1 and 2). For CO emissions, while the transport sector has the highest number of breaks, the average reduction effect size is not significant (20.8%), lower than that in the buildings sector (31.6%) (Supplementary Tables 1 and 2). This discrepancy may be due to the rebound effects of fuel economy limiting short-term benefits of fuel standard upgrades[58, 59], whereas cleaner heating fuels in buildings yield higher benefits[60, 61]. For VOCs emissions, the number of breaks in transport (21 breaks) and industrial sectors (22 breaks) are similar, but the buildings sector achieves a higher average reduction size (32.0%). The lower reduction size in the industrial sector (18.0%) may be attributed to technical challenges in detecting VOCs leaks in the petrochemical industries[62] (Supplementary Tables 1 and 2).

## 2. Attribution analysis of $O_3$ precursor emission breaks

This study employs the OECD Climate Policy Database to attribute structural breaks detected in the emission sequences of $NO_x$, CO, and VOCs. After removing redundant breaks and retaining only breaks within the same country that are not fully included in the confidence intervals of other countries, 228 distinct breaks remain. Of these, 163 are associated with the adoption or strengthening of at least one policy within a two-year window before or after the break (Fig.2, Fig.3, and Extended Data Figs. 2–5). The policy attribution method is based on 99% confidence intervals of breaks and considers expected or lagged policy effects within a ±2-year window. It incorporates known European Union (EU) multi-country policies, such as the emissions trading scheme and Minimum Energy Performance Standards for Appliances (Supplementary Table 3), as control variables to exclude confounding effects on member countries. The effectiveness of different O3 precursor emission reduction strategies varies significantly across sectors, as detailed in

Supplementary Table 4. This variation is closely linked to the policy instruments employed by different economies.

In the domain of high-frequency $NO_x$ reduction, combinations of financing mechanisms and performance standard policies have demonstrated significant synergistic effects. For instance, Chile's policy mix in 2014 resulted in a reduction in effect size of up to 32.4%. Similarly, combinations of adoption subsidies with label policies in South Africa in 2010 and adoption subsidies with performance standards in the USA in 2008 resulted in emission reductions of 46.8% and 21.9%, respectively. These results highlight the synergistic effect of policy subsidies, such as adoption subsidies and financing mechanisms, when combined with standards, including performance standards and energy efficiency labels (Fig. 4a). The ban and phase-out policy, with an average reduction effect size of 25.5%, has shown significant emission reduction effects in the electricity sector of developed countries by phasing out coal-fired power plants (Fig. 4a). Notable examples include New Zealand in 2016 (−39.4%), Spain in 2019 (−35.1%), and Greece in 2020 (−25.0%). However, it is important to emphasize that this policy needs to be applied in combination with renewable expansion planning and emission trading schemes (Supplementary Table 4). In developing countries, the renewable auction (average effect size −29.6% for the individual policy) has demonstrated even greater emission reduction potential (Fig. 4a,). Notable cases include Argentina in 2017 (−37.5%) and Brazil in 2016 (−29.6%). It is noteworthy that renewable auctions and the ban and phase-out policies, as the two most effective tools for $NO_x$ reduction, have both achieved significant pollutant reductions through structural transformation in the electricity sector (Fig. 4a).

In the domain of high-frequency CO mitigation policies, the synergistic application of adoption subsidy and carbon tax has demonstrated substantial emission reduction efficacy in the building sector of developed economies, exemplified by Canada (2005, −42.7%) and Sweden (2005, −31.5%) (Fig. 4b). Conversely, financing mechanism and fuel tax combinations exhibit pronounced advantages in developing nations' building sectors, as evidenced by Chile (2014, −51.3%) and Slovakia (2012, −46.3%) (Fig. 4b). These findings underscore the pivotal role of the building sector in achieving large-scale CO reductions, primarily through policy integration of subsidies (e.g., adoption subsidy and financing mechanism) and taxation (e.g., carbon tax and fuel tax). Notably, fuel tax demonstrates exceptional performance in developing countries (e.g., Chile, 2014, −51.3%), whereas its efficacy remains limited in developed economies (e.g., Hungary, 2009, −18.1%) (Fig. 4b). In contrast, carbon tax is predominantly adopted in developed nations but often suffers from insufficient sectoral coverage (Fig. 4b). Fossil fuel subsidy reform has been predominantly implemented in conjunction with energy efficiency mandates to achieve industrial decarbonization (e.g., Indonesia, 2013, −44.2%) (Fig. 4b and Supplementary Table 4), and paired with financing mechanisms for transport sector mitigation (e.g., Portugal, 2010, −39.5%; Indonesia, 2019, −29.8%) (Fig. 4b and Supplementary Table 4). Furthermore, the integration of public expenditure for rail and adoption subsidy policies has driven significant CO reductions in transportation systems, as illustrated by China (2009, −23.2%) and Japan (2016, −22.5%) (Fig. 4b and Supplementary Table 4).

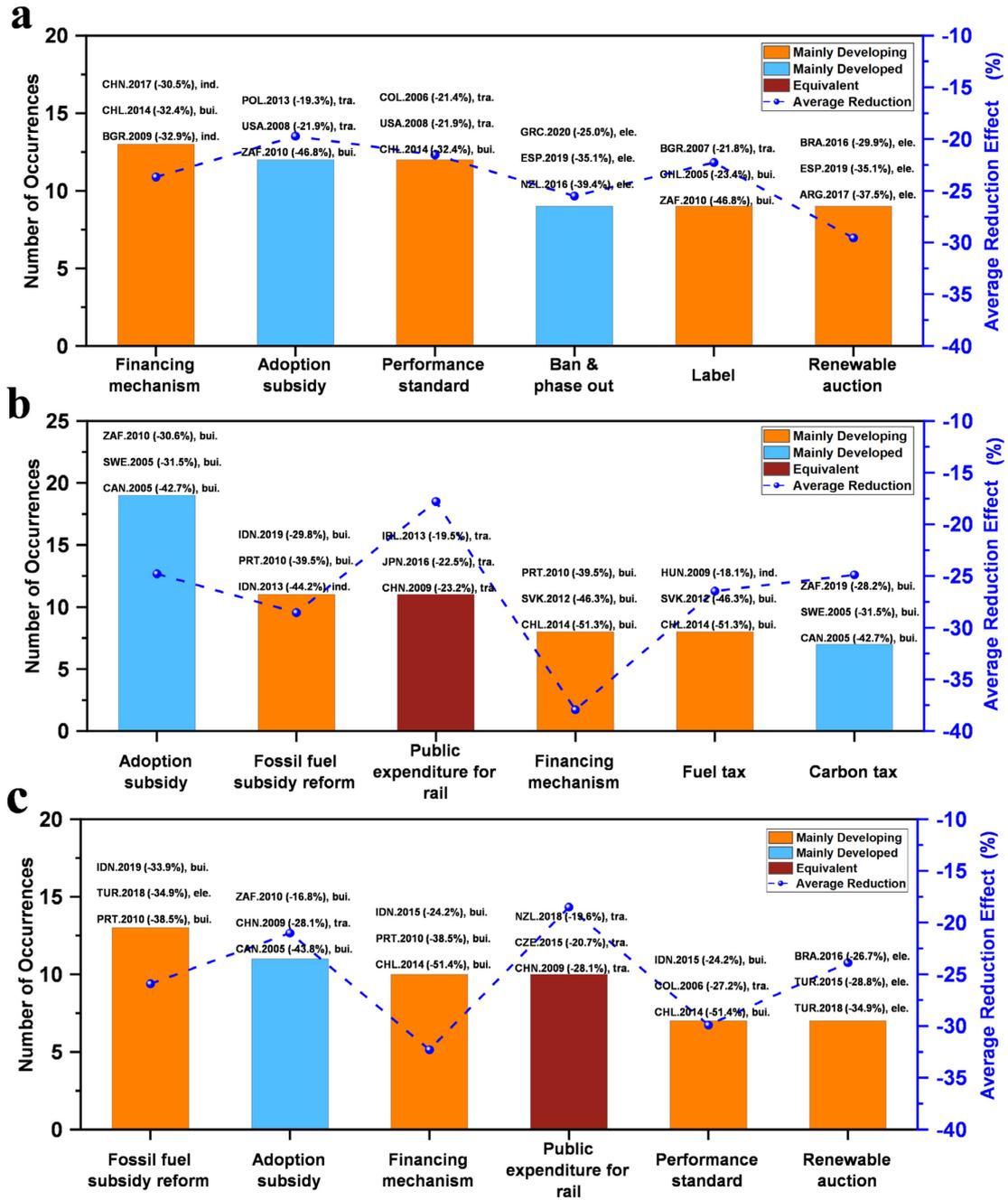

**Fig. 4 High-frequency policy analysis corresponding to emission reduction breaks for O$_3$ precursors ((a) NO$_x$, (b) CO, (c) VOCs).** The x-axis represents high-frequency policies determined through emission reduction break attribution, arranged in decreasing frequency order from left to right. The left y-axis indicates the frequency of policy matching, while the right y-axis shows the evaluated emission reduction effect. Bar color indicates predominant economic origin, mainly developed (≥67% cases), mainly developing (≥67% cases) or equivalent. The average reduction effect of these policies is depicted by blue line graphs. Top 3 representative cases are annotated on the policy bar charts with standardized nomenclature: "Policy type: Country. Year (effect size %), sector code". For instance, "Financing mechanism: Bulgaria. 2009 (-32.9%), ind." denotes a financing mechanism policy implementation in Bulgaria (2009) achieving significant emission reduction effect size (-32.9%) in the industrial sector (with "bui." = building sector, "tra." = transport sector, and "ele." = electricity sector). These data are sourced from the Extended Data Table. 1.
13

In the context of high-frequency policies for VOCs abatement, fossil fuel subsidy reform demonstrates an average effect size of −25.9% (Fig. 4c). When implemented in conjunction with financing mechanisms, its mitigation potential significantly increases, as evidenced by case studies in Portugal (2010, −33.9%) and Indonesia (2019, −38.5%) (Fig. 4c and Supplementary Table 4). Public expenditure for rail exhibits a lower average effect size (−18.5%), but achieves substantial VOCs reductions in the transport sector when combined with adoption subsidies, exemplified by China (2009), Czech Republic (2015), and New Zealand (2018) (Fig. 4c and Supplementary Table 4). Performance standards are frequently integrated with financing mechanisms to enhance emission reductions, particularly in the building sectors of developing economies, as observed in Chile (2014, −51.4%) and Indonesia (2015, −24.2%) (Fig. 4c). Renewable auctions show pronounced efficacy in the electricity sectors of developing nations, especially when combined with fossil fuel subsidy reform and renewable expansion planning, as demonstrated by Turkey's policy outcomes in 2018 (−34.9%) and 2015 (−28.8%) (Fig. 4c and Supplementary Table 4). These results align with findings from the International Institute for Sustainable Development, which highlight Turkey's success in reducing VOCs emissions through coal subsidy reforms and increased renewable energy investments[63].

## 3. Identification of effective policy mixes

To systematically compare the emission reduction effects of 233 policy instruments matched with emissions breaks, whether implemented individually and in combination, we conduct an integrated policy evaluation. Tools that share similar economic mechanisms (e.g., emission trading scheme, carbon taxes, and fuel taxes, hereafter collectively referred to as taxation) are classified as one category. Fig. 5 summarizes policies aimed at reducing emissions of $NO_x$, CO, and VOCs, comparing the average emission reductions achieved through individual and joint implementations, and further examining the synergistic effects when non-price policies (i.e. those excluding fossil fuel subsidy reforms and taxation) paired with pricing tools (fossil fuel subsidy reforms and taxation).

In most cases, the joint implementation of policy instruments yields significantly greater effects than their individual application (Fig. 5). For example, labeling schemes (in the transport sector), performance standards in the buildings sector, and transportation labeling as well as fossil fuel subsidy reforms can only achieve significant emission reductions when implemented as part of a combination, indicating limited individual impact or low frequency of adoption alone. The combination of tools such as bans and phase out, building codes, subsidy measures, and energy efficiency requirements has a markedly better effect (Fig. 5). Notably, when combining financial mechanisms with pricing tools, the reduction size was -28.0%, far exceeding the -17.9% when implemented separately (Fig. 5). In the industrial sector, the joint application of energy efficiency standards resulted in an average reduction size of -25.1%, compared to -15.8% reduction for individual implementation (Fig. 5). Taxation remains the only instrument that, if implemented separately, achieve comparable or even greater effects across all sectors, particularly in the electricity and industrial sectors where pricing is included in all effective investment portfolios. When non-price policies are paired with pricing tools, their synergistic benefits are evident. Given the limitations of individual data points and the pronounced heterogeneity in portfolio synergy, policy design must be tailored to specific economic backgrounds and sectoral characteristics.

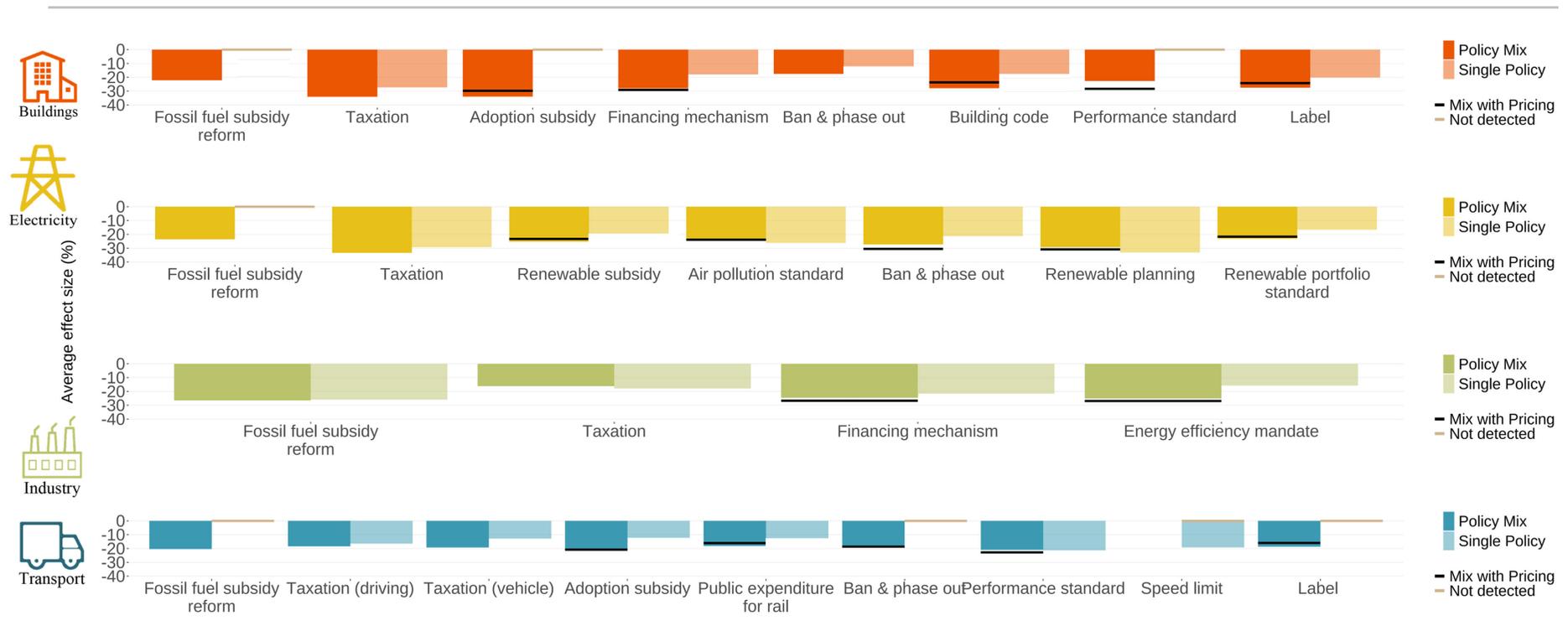

**Fig. 5. Effective policies and policy mixes for O$_3$ precursors (NOx, CO, VOCs).** On the basis of point estimates for emission breaks in specific countries (Supplementary Table 4), we compared the average effect sizes of all breaks in which a policy instrument appears individually with those of all breaks in which this policy instrument appears in a mix. For non-price-based policies, the thick black line also indicates the average effect size of a mix with a given policy instrument and pricing (through taxation or reducing fossil fuel subsidies).

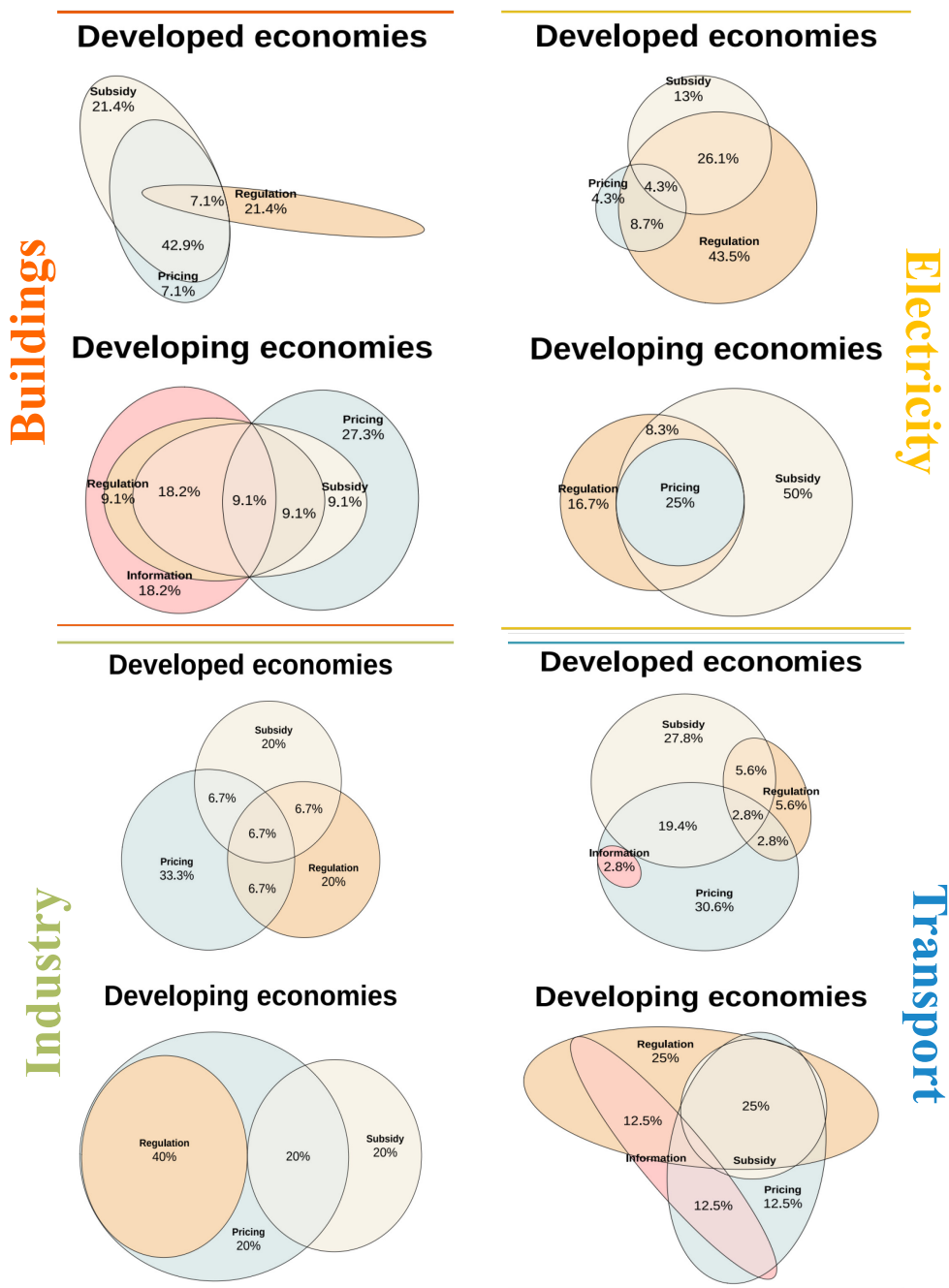

**Fig. 6 The Euler diagrams display the combinations of effective policy types for various sectors of O₃ precursors (NO$_x$, CO, VOCs) in both developed and developing economies.** For each circular area, the percentages indicate which share of successful interventions in this sector is composed of a specific individual policy type or a specific combination of policy types. A single policy type encompasses breaks that match a single policy instrument (such as a pricing scheme) or a combination of policy tools of the same type (such as two or more different pricing schemes).

In addition, we divide policy tools into four categories (Supplementary Table 5), which are the core of high-level political discussions on optimal policy design: information, pricing, regulation, and subsidy. Fig. 6 presents the proportion of emission reductions achieved by each unique policy combination or independent policy among all successful interventions. The transport sector exhibits the strongest complementarity: Pricing emerges as the most complementary instrument, with 60% of policy portfolios in developing economies relying on pricing combinations, while developed economies predominantly use independent pricing (31%) and subsidies (28%) (Fig. 6, transport). In the buildings sector, there are divergent strategies: 83% of portfolios in developing economies are driven by regulatory portfolios, while developed economies tend to adopt a mixed subsidy-pricing strategy (50%) (Fig. 6, buildings). The electricity sector is mainly dominated by independent tools, with 44% of emission reductions in developing economies stemming from regulatory policies, while 50% of emissions reductions in developed economies come from subsidies (Fig. 6, electricity). In the industrial sector, the trends are polarized: developed economies prefer independent pricing (33%), whereas 60% of emission reduction policies in developing economies depend on pricing portfolios (Fig. 6, industry). These findings underscore the necessity of customized policy design methods with pricing as the synergistic core (Fig. 6).

## 4. Robustness Checks and Model Validation

To assess the robustness and reliability of our model, we conduct several sensitivity analyses. Firstly, we employ the Impulse Indicator Saturation (IIS) method to examine the sensitivity of our break detection algorithm. By introducing 0/1 dummy variables to control for the influence of outliers, we evaluat the model's robustness to potential data mis-specifications[64, 65]. The results indicate that the number and variability of retained IIS metrics are stable, indicating that our baseline model is appropriately defined and not unduly influenced by outliers (Supplementary Tables 7 -30). Secondly, we adjust the false positive rate threshold and expand the target p-value from 0.1% to 1%. This adjustment allows us to test the stability of our main model results at both stricter and looser levels. The findings show that the results remain consistent within these different thresholds, further confirming the robustness of our detection process (Supplementary Tables 7 -30). Thirdly, after identifying significant emission reduction breaks using our panel model, we apply the Generalized Synthetic Control Method (GSCM) to validate the estimated effect sizes. GSCM constructs a synthetic control unit by reweighting untreated countries to closely match the economic characteristics and emission trends of the treated unit during the pretreatment period[66]. This method accounts for unobserved time-varying confounders, enhancing the comparability between treated and control units. The GSCM results are highly consistent with our main findings, strengthening the credibility of our causal inferences.

It's worth noting that GSCM requires a sufficient number of pre-treatment periods (typically at least five) to produce reliable estimates. Consequently, GSCM cannot provide effective estimates for breaks detected before 2005. Therefore, our main text emphasizes the TWFE results, while GSCM findings are presented in the supplementary as part of our sensitivity analysis. Supplementary Figs. 1 through 24 provide detailed visualizations of policy interventions and their temporal distributions across different sectors and countries. These figures display both the pre-treatment fit and the estimated policy effects, and compare counterfactual predictions with actual observations. The consistency between the TWFE and GSCM method in different specifications and robustness checks cross-validates our conclusions and underscores the reliability of our model (Supplementary Table 32 and Figs. 1 through 24). By integrating multiple methodologies and conducting thorough robustness checks, we conduct a comprehensive and validated assessment of the causal impacts of environmental policies in diverse national contexts.

# Discussion

Identifying effective policy interventions is crucial for guiding policymakers in designing impactful measures. Based on nearly two decades of global climate policy data and $O_3$ precursor emission records, we apply an EDTWFE-DID model alongside a machine-learning-driven structural break detection method to identified 78, 77 and 78 structural breaks in $NO_x$, CO and VOCs emissions, respectively, across the building, electricity, industrial, and transport sectors in 41 countries. From these breaks, we estimate equivalent total emission reductions of 0.96–0.97 Gt for $NO_x$, 2.84–2.88 Gt for CO and 0.47–0.48 Gt for VOCs. We observe significant differences in the emission reduction policies and effectiveness of various $O_3$ precursors across different sectors and economies (Supplementary Table 4). For successful $NO_x$ mitigation, structural transformation-driven policies in the electricity sector have played a pivotal role. In developed economies, the combined implementation of ban & phase-out policies and emission trading schemes has effectively accelerated the closure or retrofitting of high-emission power plants, achieving a reduction of up to 32.4%. This is exemplified by the U.S. Acid Rain Program, which established emission caps and enabled tradable permits, resulting in significant $NO_x$ abatement in the electricity sector[67]. In developing nations, renewable energy auctions have proven instrumental (-29.6% mean efficacy), as evidenced by the International Renewable Energy Agency report highlighting cost reductions in renewable electricity procurement through competitive auctions in countries such as Brazil[68]. For volatile organic compounds (VOCs), fossil fuel subsidy reform paired with financing mechanisms demonstrates a reduction of up to 38.5%. Additionally, the combination of fossil fuel subsidy reform, renewable auctions, and renewable expansion planning yields notable results (reduction of up to 34.9%) in developing economies' electricity sectors. In carbon monoxide (CO) mitigation, the building sector emerges as a critical domain. Developed economies achieve significant CO reductions through the adoption of subsidies and carbon tax integration, while developing nations rely on financing mechanisms combined with fuel taxes. These findings align with International Energy Agency analyses identifying fiscal incentives—including grants, concessional loans, and tax relief—as key drivers of decarbonization and climate action in the building sector[69].

We find that the combination of non-price-based policies, especially financial mechanisms, adoption subsidy, ban & phase out, performance standard, labeling, energy efficiency mandate, combined with pricing tools, can result in significant synergistic effects, and greatly improve overall emission reduction outcomes. This finding is consistent with previous theoretical and empirical studies on policy complementarities[32, 70], indicating that well-designed policy combinations can address the limitations of individual policies, alleviate market failures, and enhance the credibility and rigor of environmental policies[71, 72]. In addition, differences in policy instruments preferences across sectors and economic development stages reflect varying institutional capacities, market structures, and policy objectives[73, 74]. Our cross-sector analysis of policy instrument preferences specifically reveals distinct dependencies across industries and economic development stages. In the transport sector, pricing instruments dominate in both developing and developed economies, demonstrating strong complementarity. The building sector shows marked divergence: regulatory combinations prevail in developing economies, while developed economies rely on pricing-subsidy hybrid strategies. The electricity sector is primarily driven by single-policy approaches, with developing economies favoring regulatory tools and developed economies depending on subsidies.

Industrial decarbonization exhibits a polarized pattern: developed economies prioritize standalone pricing, whereas developing economies adopt pricing combinations.

Methodologically, our study advances the use of machine learning to enhance structural break detection, overcoming the limitations of traditional TWFE models that require the pre-specified policy intervention points. This approach allows for flexible identification of unknown policy effects, and our robustness checks with generalized synthetic control methods confirm the reliability of our causal inferences. However, our analysis is limited by regional imbalances, primarily due to the scarcity of policy data in developing countries, particularly in Africa and Asia. Future research should aim to expand the sample coverage and incorporate more dynamic indicators to more comprehensively evaluate policy synergies.

In summary, our empirical results demonstrate the effectiveness of combining pricing tools with non-price-based policies, providing robust empirical evidence for building a comprehensive policy toolkit centered on pricing mechanisms, supplemented by regulation, subsidies, and information incentives. Policymakers should develop emission reduction strategies that are both universally applicable and tailored to the specific technological characteristics and market conditions of each sector. However, achieving larger-scale emission reductions should also consider factors beyond environmental benefits and address broader policy issues.

# Extended Data Figure

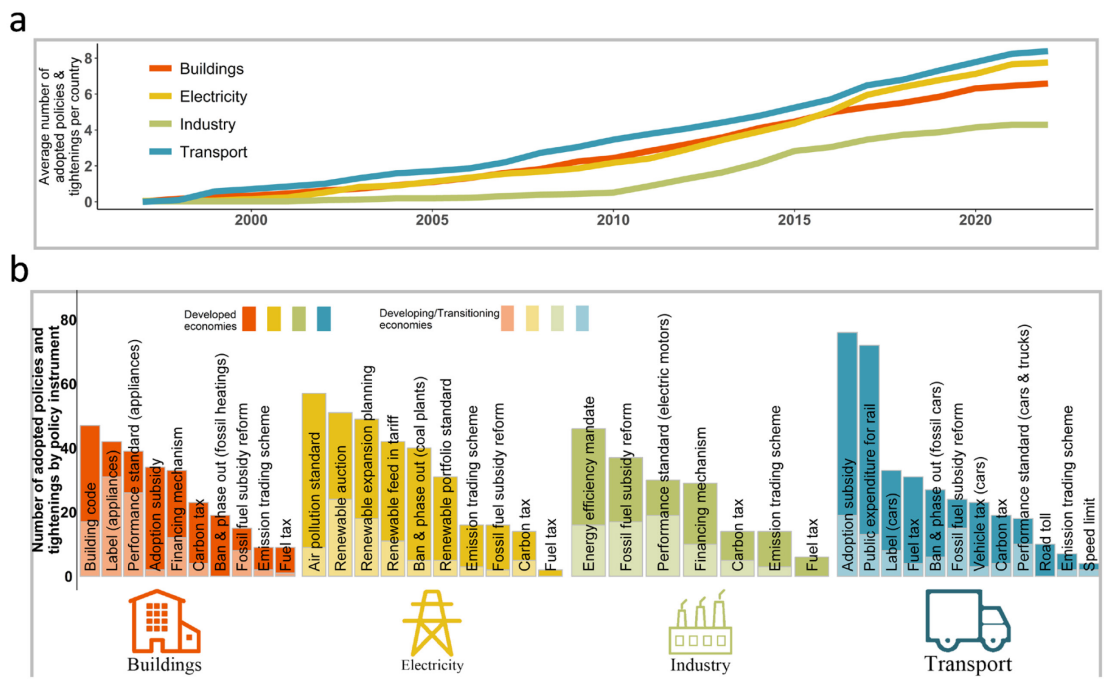

**Extended Data Fig. 1 | Characterization of climate policy frameworks.** (a) Temporal evolution of cross-sectoral climate policies (1998–2022), showing sustained growth across economic sectors based on OECD policy inventory data. (b) Sectoral distribution of policy instrument typologies, contrasting deployment patterns between developed and developing/transitioning economies. Command-and-control measures (e.g., emission standards, technology mandates) dominate all sectors except transport, accounting for 270 measures cumulatively. Market-based instruments (subsidies, carbon pricing) show concentrated adoption in developed economies, particularly within transport. Notably, while subsidies (116 measures) dominate market-based instruments, carbon pricing mechanisms (carbon taxes and emission trading schemes) remain limited, with 88% implemented in developed economies.

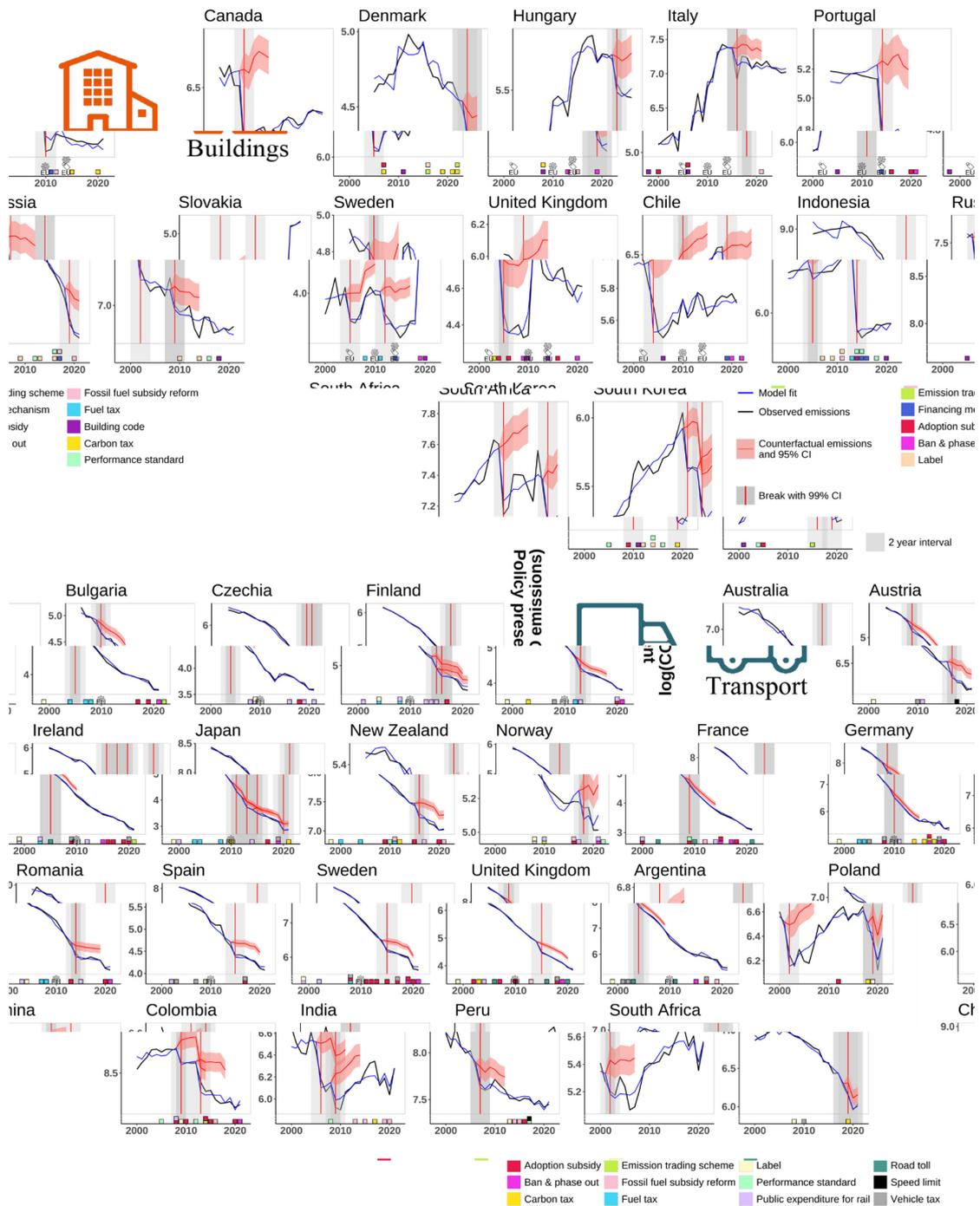

**Extended Data Fig. 2 | Breaks detection and policy matching for CO emissions in the buildings and transport sectors.**

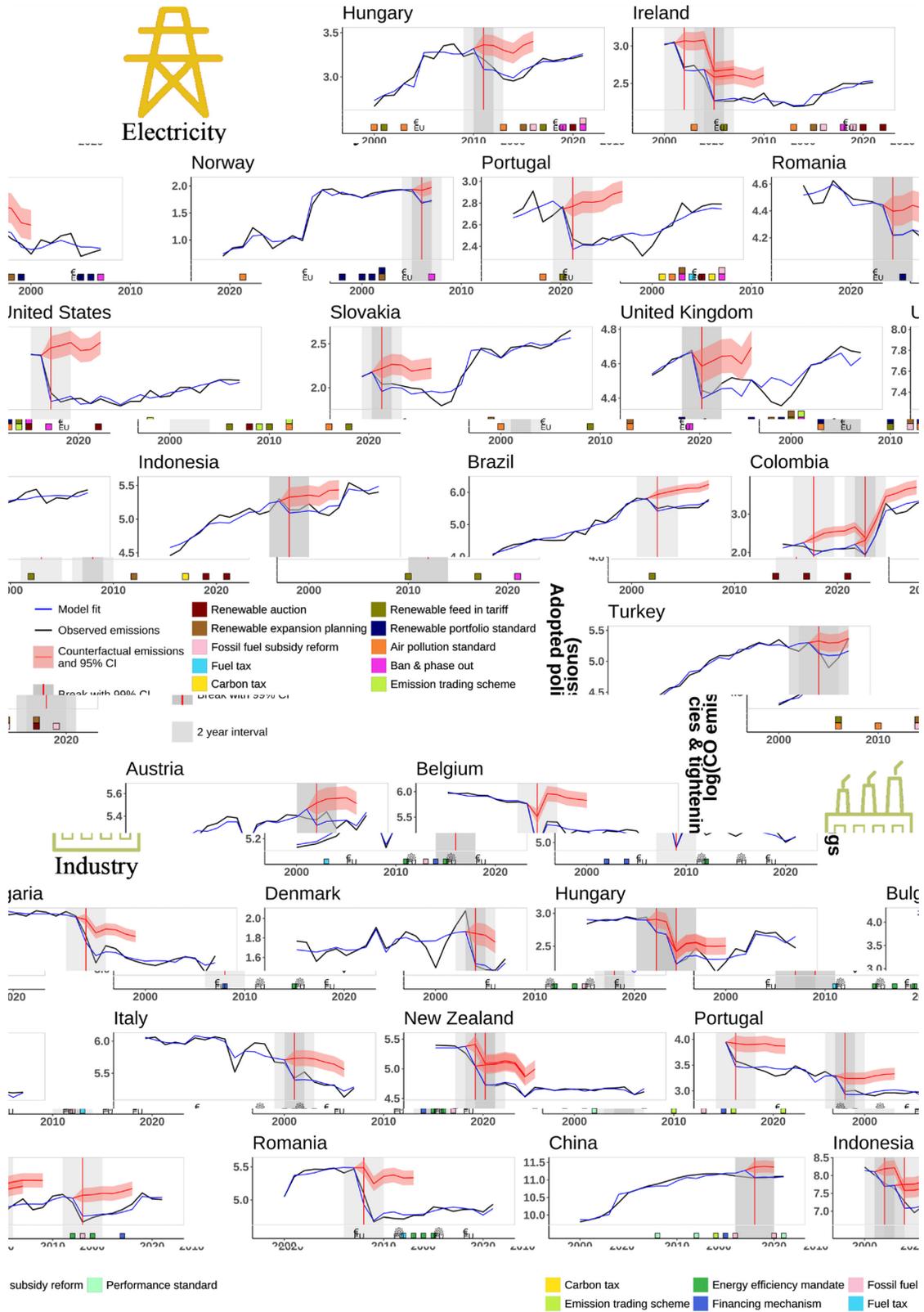

**Extended Data Fig. 3 | Breaks detection and policy matching for CO emissions in the electricity and industrial sectors.**

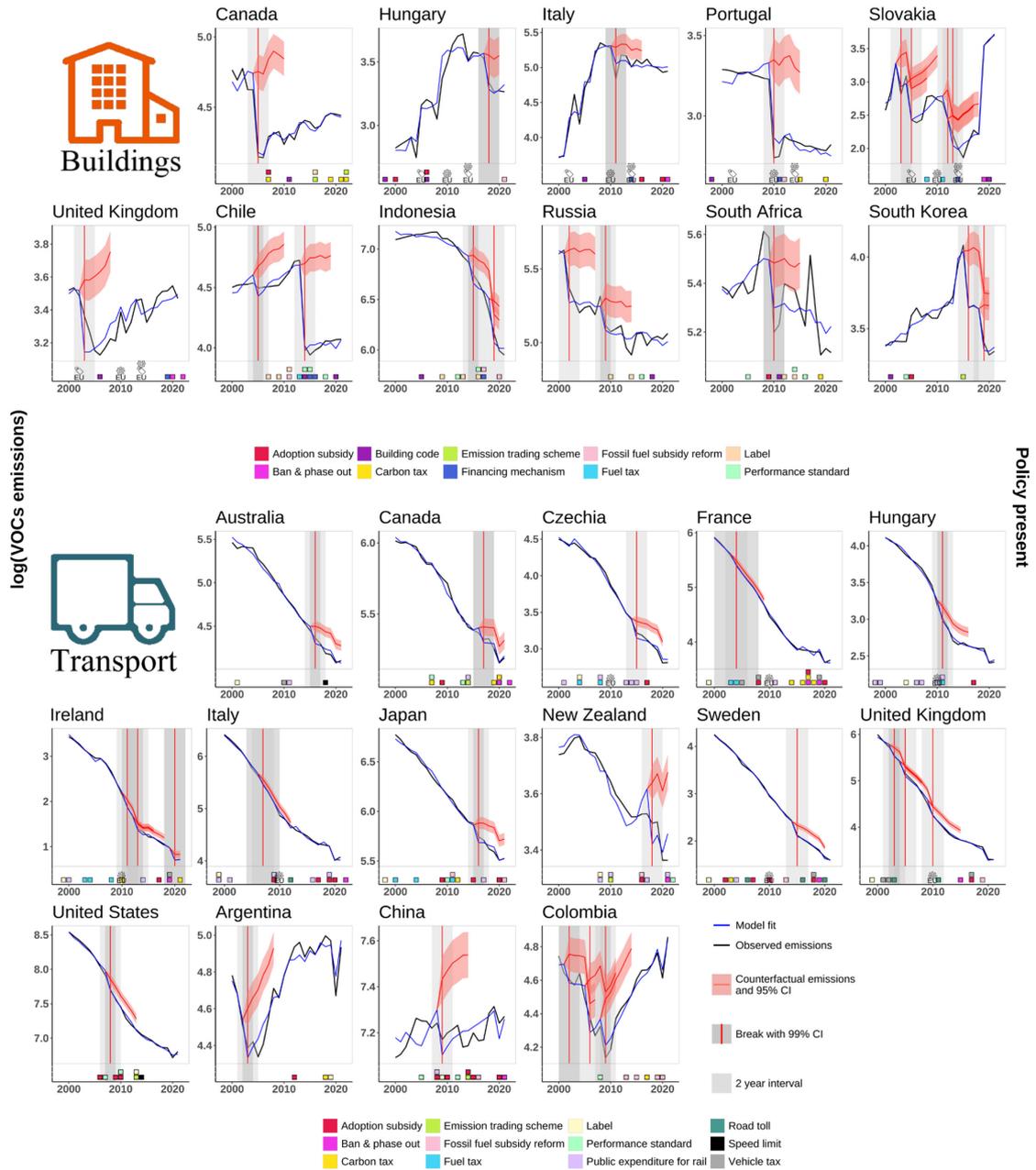

**Extended Data Fig. 4 | Breaks detection and policy matching for VOCs emissions in the buildings and transport sectors.**

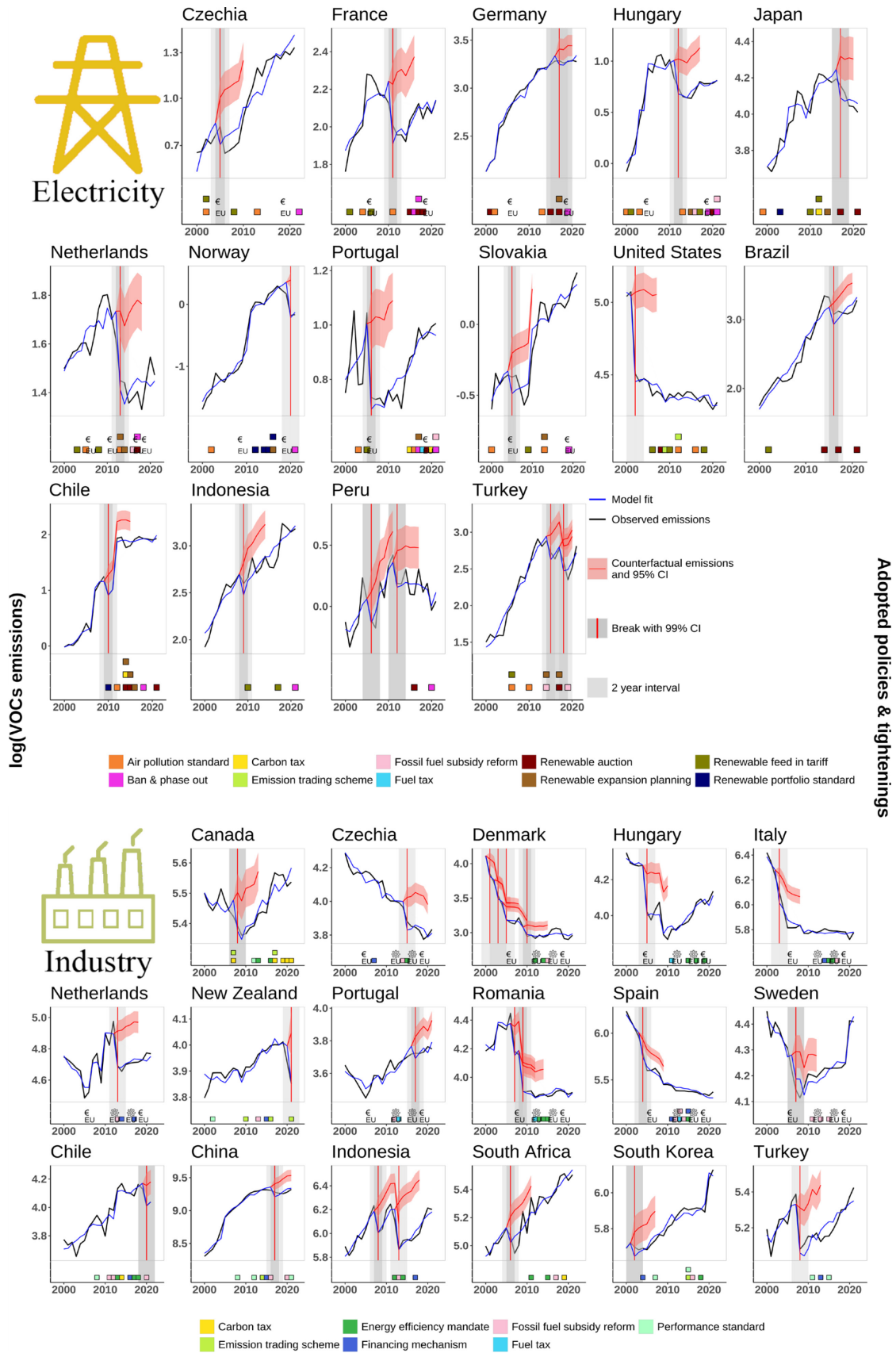

**Extended Data Fig. 5 | Breaks detection and policy matching for VOCs emissions in the electricity and industrial sectors.**

**Extended Data Table. 1 | High frequency policy data on emission reduction of O$_3$ precursors (NO$_x$, CO, VOCs)**. These data are sourced from the original Supplementary Table 4, which calculates the frequency of occurrence and average emission reduction effect of each policy. The top six policies are listed on the basis of their application frequency, and specific cases under these high-frequency policies are analyzed by selecting the top three cases with the highest absolute emission reduction effects from the original data.

| NO$_x$ Policy Designation | Frequency | Average Reduction Effect | Economic Typology Distribution | Exemplary Cases |
| --- | --- | --- | --- | --- |
| Financing mechanism | 13 | -0.2367 | Developing dominated (n=9) | Bulgaria.2009 (-32.9%) in industry, Chile.2014 (-32.4%) in buildings, China.2017 (-30.5%) in industry |
| Adoption subsidy | 12 | -0.1973 | Developed dominant (n=9) | SouthAfrica.2010 (-46.8%) in buildings, UnitedStates.2008 (-21.9%) in transport, Poland.2013 (-19.3%) in transport |
| Performance standard | 12 | -0.2153 | Developing dominated (n=8) | Chile.2014 (-32.4%) in buildings, UnitedStates.2008 (-21.9%) in transport, Colombia.2006 (-21.4%) in transport |
| Ban & phase out | 9 | -0.2551 | Developed dominant (n=7) | NewZealand.2016 (-39.4%) in electricity, Spain.2019 (-35.1%) in electricity, Greece.2020 (-25.0%) in electricity |
| Label | 9 | -0.2227 | Developing dominated (n=7) | SouthAfrica.2010 (-46.8%) in buildings, Chile.2005 (-23.4%) in buildings, Bulgaria.2007 (-21.8%) in transport |
| Renewable auction | 9 | -0.2956 | Developing dominated (n=6) | Argentina.2017 (-37.5%) in electricity, Spain.2019 (-35.1%) in electricity, Brazil.2016 (-29.9%) in electricity |
| CO Policy Designation | Frequency | Average Reduction Effect | Economic Typology Distribution | Exemplary Cases |
| Adoption subsidy | 19 | -0.2479 | Developed dominant (n=15) | Canada.2005 (-42.7%) in buildings, Sweden.2005 (-31.5%) in buildings, South Africa.2010 (-30.6%) in buildings |
| Fossil fuel subsidy reform | 11 | -0.2854 | Developing dominated (n=9) | Indonesia.2013 (-44.2%) in industry, Portugal.2010 (-39.5%) in buildings, Indonesia.2019 (-29.8%) in buildings |
| Public expenditure for rail | 11 | -0.1779 | Equivalent | China.2009 (-23.2%) in transport, Japan.2016 (-22.5%) in transport, Ireland.2013 (-19.5%) in transport |
| Financing mechanism | 8 | -0.3794 | Developing dominated (n=6) | Chile.2014 (-51.3%) in buildings, Slovakia.2012 (-46.3%) in buildings, Portugal.2010 (-39.5%) in buildings |
| Fuel tax | 8 | -0.2646 | Developing dominated (n=6) | Chile.2014 (-51.3%) in buildings, Slovakia.2012 (-46.3%) in buildings, |

| | | | | Hungary.2009 (-18.1%) in industry |
|---|---|---|---|---|
| Carbon tax | 7 | -0.2488 | Developed dominant (n=5) | Canada.2005 (-42.7%) in buildings, Sweden.2005 (-31. 5%) in buildings, South Africa.2019 (-28.2%) in buildings |
| VOCs Policy Designation | Frequency | Average Reduction Effect | Economic Typology Distribution | Exemplary Cases |
| Fossil fuel subsidy reform | 13 | -0.2592 | Developing dominated (n=10) | Portugal.2010 (-38.5%) in buildings, Turkey.2018 (-34.9%) in electricity, Indonesia.2019 (-33.9%) in buildings |
| Adoption subsidy | 11 | -0.2101 | Developed dominant (n=8) | Canada.2005 (-43.8%) in buildings, China.2009 (-28.1%) in transport, South Africa.2010 (-16.8%) in buildings |
| Financing mechanism | 10 | -0.3229 | Developing dominated (n=7) | Chile.2014 (-51.4%) in buildings, Portugal.2010 (-38.5%) in buildings, Indonesia.2015 (-24.2%) in buildings |
| Public expenditure for rail | 10 | -0.1851 | Equivalent | China.2009 (-28.1%) in transport, Czech Republic.2015 (-20.7%) in transport, New Zealand.2018 (-19.6%) in transport |
| Performance standard | 7 | -0.2991 | Developing dominated (n=5) | Chile.2014 (-51.4%) in buildings, Colombia.2006 (-27.2%) in transport, Indonesia.2015 (-24.2%) in buildings |
| Renewable auction | 7 | -0.2388 | Developing dominated (n=6) | Turkey.2018 (-34.9%) in electricity, Brazil.2016 (-26.7%) in electricity, Turkey.2015 (-28.8%) in electricity |